%% file: NYMGastro2020.tex
\documentclass[12pt]{article}
\usepackage{graphics,graphicx}
\usepackage{times}
\usepackage{geometry}
\usepackage[utf8]{inputenc}
\usepackage{enumitem,amssymb}
\usepackage{ragged2e}
\newlist{thematic}{itemize}{8}
\setlist[thematic]{label=$\square$}
\usepackage{pifont}

\usepackage[numbers]{natbib}
\usepackage[colorlinks=true,
            linkcolor=blue,
            urlcolor=blue,
            citecolor=blue]{hyperref}

\newcommand\aj{{AJ}}
\newcommand\araa{{ARA\&A}}
\newcommand\apj{{ApJ}}
\newcommand\apjl{{ApJL}}     
\newcommand\aap{{A\&A}}
\newcommand\mnras{{MNRAS}}

\begin{document}
\raggedright
\huge
Astro2020 Science White Paper \linebreak

The Early Evolution of Stars and Exoplanet Systems: Exploring and Exploiting Nearby, Young Stars \linebreak
\normalsize

\noindent \textbf{Thematic Areas:} \hspace*{60pt} $\square$ Planetary Systems \hspace*{10pt} (1) Star and Planet Formation \hspace*{20pt}\linebreak
$\square$ Formation and Evolution of Compact Objects \hspace*{31pt} $\square$ Cosmology and Fundamental Physics \linebreak
  (2)  Stars and Stellar Evolution \hspace*{1pt} $\square$ Resolved Stellar Populations and their Environments \hspace*{40pt} \linebreak
  $\square$    Galaxy Evolution   \hspace*{45pt} $\square$             Multi-Messenger Astronomy and Astrophysics \hspace*{65pt} \linebreak
  
\textbf{Principal Author:}

Name: Joel H. Kastner	
 \linebreak						
Institution: Rochester Institute of Technology, Rochester, NY, USA 
 \linebreak
Email: jhk@cis.rit.edu
 \linebreak
Phone: (585) 475-7179 
 \linebreak
 
\textbf{Co-authors:}  \\
Katelyn Allers (Bucknell University, Lewisburg, PA, USA); Brendan Bowler (U. Texas, Austin, TX, USA); Thayne Currie (NASA ARC/NAOJ Subaru Telescope, Hilo, HI, USA); Jeremy Drake (SAO, Cambridge, MA, USA), Trent Dupuy (Gemini Observatory, Hilo, HI, USA); Jackie Faherty (AMNH, New York, NY, USA); Jonathan Gagn\'{e} (U. Montreal, PQ, Canada); Michael Liu (U. Hawaii, Honolulu, HI, USA); Eric Mamajek (NASA/JPL, Pasadena, CA, USA); Dimitri Mawet (CalTech, Pasadena, CA, USA); Evgenya Shkolnik (Arizona State U., Tempe, AZ, USA); Inseok Song (U. Georgia, Athens, GA, USA); Russel White (Georgia St.\ U., Atlanta, GA, USA); Ben Zuckerman (UCLA, Los Angeles, CA, USA)
  \linebreak

\textbf{Abstract:} 
Our knowledge of the population of young (age $\stackrel{<}{\sim}$750 Myr) stars that lie within $\sim$120 pc of the Sun is rapidly accelerating. The vast majority of these nearby, young stars can be placed in kinematically coherent groups (nearby, young moving groups; NYMGs). NYMGs and their member stars afford unmatched opportunities to explore a wide variety of aspects of the early evolution of stars and exoplanet systems, including stellar initial mass functions and age determination methods; the magnetic activities and high-energy radiation environments of young, late-type stars; the dynamics of young binary and hierarchical multiple systems; the late evolutionary stages of circumstellar disks; and, especially, direct-imaging discovery and characterization of massive young exoplanets. In this White Paper, we describe how our understanding of these and many other aspects of the early lives of stars and planetary systems is ripe for progress over the next decade via the identification and study of NYMG members with present and next-generation facilities and instruments.

\pagebreak

\setcounter{page}{1}

\input{NYMGastro2020text}

\pagebreak

\input{RefList2.tex}

\end{document}

%% file: NYMGastro2020text.tex
\section{A New, Rapidly Accelerating Field of Astrophysics}
\vspace{-.08in}

Astronomers' knowledge of the population of stars with ages less than $\sim$750 Myr that lie within $\sim$120 pc of the Sun has accelerated dramatically over the past two decades \cite{Kastner2016IAUSed}. The vast majority of these nearby, young stars can be placed in kinematically coherent, approximately coeval groups (typically referred to as nearby, young moving groups, hereafter NYMGs; e.g., \cite{Mamajek2016}). Various efforts are now underway to exploit the initial data releases from the {\it Gaia} space astrometry mission \cite{Gaia2018} to study known NYMG members, confirm suspected NYMG members, and identify new candidate NYMG members (e.g., \cite{Faherty2018,Gagne2018a, Gagne2018DR1, Gagne2018BAN,LeeSong2018,LeeSong2019,Zuckerman2019}). As a consequence, the number of candidate members of NYMGs has exploded, from a few hundred circa 2008 \cite{ZuckermanSong2004,Torres2008} to well over a thousand, and counting, as of early 2019 (Fig.~1; \cite{Gagne2018a}).

\begin{figure*}[h!]
\centering
\includegraphics[width=6in,angle=0]{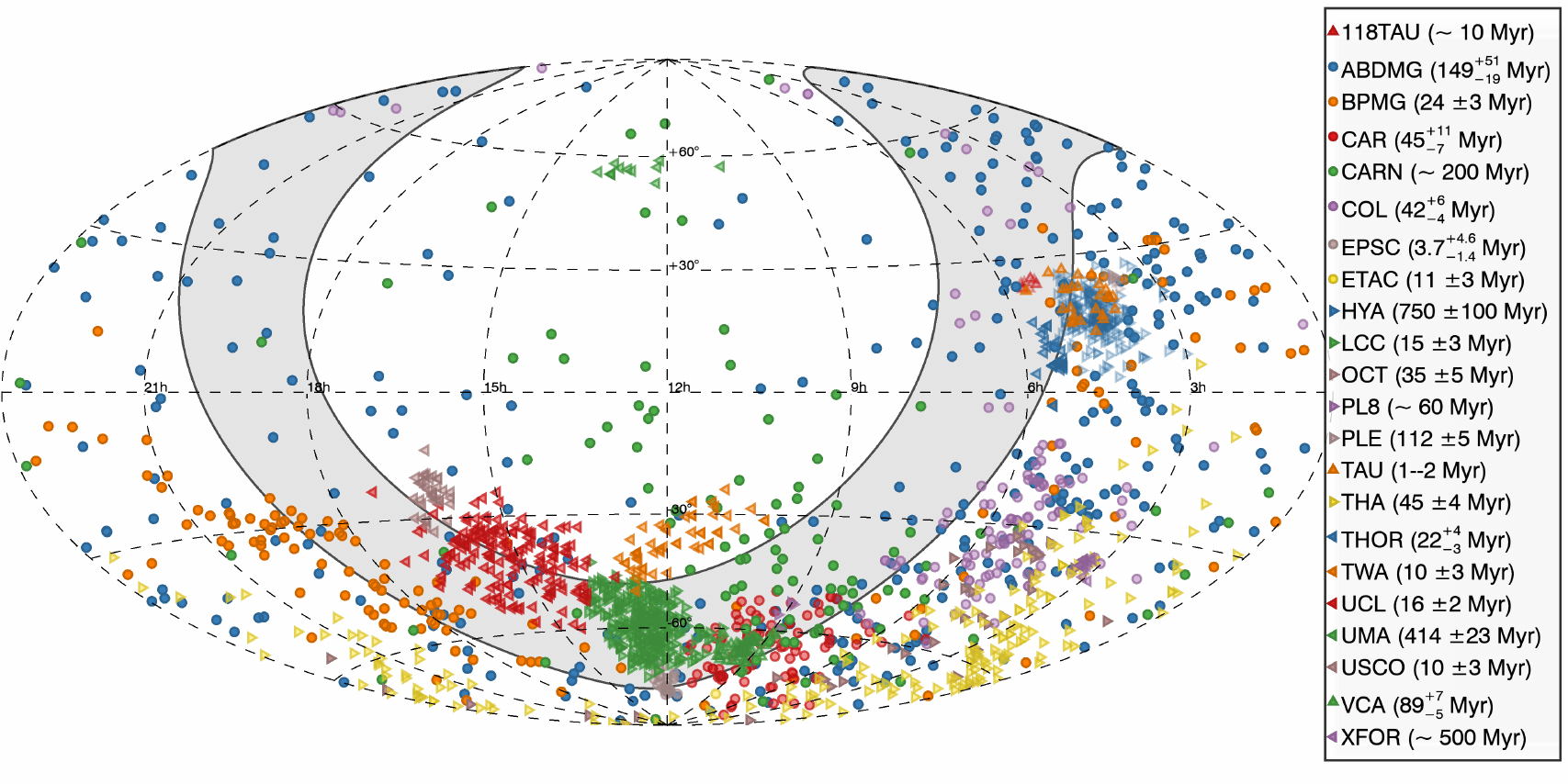}
\vspace{-.1in}
\caption{\it \small Sky distribution of known and candidate NYMG members within 120 pc, as of early 2019 (J. Gagne, pvt.\ comm.). The grey band indicates the galactic plane ($\pm15^\circ$). Key to MG assignments \cite{Gagne2018BAN}: 118TAU = 118 Tau Assoc.; ABDMG = AB Dor MG; BPMG = $\beta$ Pic MG; CAR = Carina; CARN = Carina Near; COL = Columba; EPSC = $\epsilon$ Cha Assoc.; ETAC = $\eta$ Cha cluster; HYA = Hyades; LCC = Lower Crux Cen; OCT = Octans MG; PL8 = Platais 8; PLE = Pleiades; TAU = Taurus; THA = Tuc Hor Assoc.; THOR = 32 Ori Assoc.; TWA = TW Hya Assoc.; UCL = Upper Cen Lup; UMA =Ursa Major; USCO =  Upper Sco Assoc.; VCA = Volans Car Assoc.; XFOR = $\chi^1$ For Assoc.}
\label{fig:GagneDR2}
\end{figure*}

NYMGs and their member stars afford unmatched opportunities to explore a wide variety of aspects of the early evolution of stars and exoplanet systems. Foremost among these opportunities is the potential to carry out direct imaging searches for, and followup spectroscopy of, young exoplanets; indeed, {\it nearby young stars are uniquely well-suited to direct-imaging exoplanet studies} (\S 2.3).  Other areas that are especially ripe for progress over the next decade via the exploration and exploitation of young stars near the Sun include determination of the present-day initial mass function in the solar neighborhood, the statistics and dynamics of young binary and hierarchical multiple systems, the magnetic activities and high-energy radiation environments of young late-type stars, and the late planet-building stages and dispersal of circumstellar, protoplanetary disks. 

\vspace{-.08in}
\section{Exploring and Exploiting Nearby, Young Stars}
\vspace{-0.08in}

As of April 2018, the Gaia mission had already provided trigonometric parallax distances and proper motions for more than a billion stars at unprecedented precision, as well as heliocentric radial velocities (RVs) for a million of these stars. The mission is scheduled for additional data releases in 2020, 2021 and 2022. The resulting avalanche of Galactic positions ($XYZ$) and kinematics ($UVW$ velocities) for stars in the solar neighborhood will dramatically improve our understanding of the roughly two dozen NYMGs within 120 pc of the Sun (Fig.~1). We have currently only seen the tip of the iceberg of these young associations in terms of their membership, because faint low-mass stars, for which we were previously badly lacking detailed kinematic data, dominate their stellar populations. As a consequence of the superb astrometric and photometric quality and sheer volume of Gaia data 
and the parallel advance of sophisticated membership survey and analysis tools \cite{LeeSong2018,Gagne2018a,Crundall2019}, we expect that nearly all members of NYMGs that lie within $\sim$120 pc will be discovered within the next few years. The list of NYMG members will reach down to the low mass end of brown dwarfs (including free-floating massive planets, for the nearest young moving groups \cite{Gagne2018RNAAS,Faherty2019}), and the total number of members (excluding the rich Sco-Cen sub-groups) will approach $\sim$3,000. This near-complete membership census will have cascading impacts on a broad range of problems in the early evolution of stars and planetary systems, enabling new approaches to longstanding problems as well as entirely new directions of scientific inquiry. 

\input{Table}

Below and in Table 1, we summarize various interwoven lines of investigation enabled by the study of NYMGs and their members, and the key advances necessary to make significant progress in each of these areas over the coming decade\footnote{Access to the southern hemisphere is essential for the study of NYMGs (Fig.~1) so it is vital that the U.S. community retain access to small, large, and extremely large optical/IR telescopes, as well as radio interferometers, south of the equator.}. These investigations can be loosely grouped into three main themes: young star population studies; the astrophysics of pre-main sequence and young main sequence stars; and direct-imaging detection and characterization of exoplanets and exoplanet formation environments. 

\vspace{-.08in}
\subsection{Young star populations}
\vspace{-.08in}
{\bf Mass functions and multiplicity.} Complete mass functions for stars in the age range represented by the known NYMGs within $\sim$120 pc --- i.e., $\sim$10--750 Myr --- have been notoriously difficult to obtain, because of the lack of nearby star clusters spanning this range. In addition to completing the membership census of NYMG members (see preceding), Gaia data will yield the visual binary and companion frequencies of these nearby young associations for separations from tens to thousands of au \cite{ElBadry2018}. These results, combined with comprehensive adaptive optics imaging surveys and high-precision RV surveys of NYMG members (\S 3), will specify to unprecedented accuracy the present-day, solar-neighborhood initial mass function for the poorly constrained regime below $\sim$1 $M_\odot$. We can furthermore extend our knowledge of the binary and multiplicity fractions for stars of age $\sim$10--750 Myr down to mass ratios $\sim$0.01, with profound implications for the statistics of young binary systems \cite{Oh2017,Malkov2018} and our understanding of the stability and dissolution of hierarchical multiple systems \cite{Kastner2018RNAAS}. \\
{\bf NYMGs: merging, splitting, discovering, and discarding.} We must take advantage of the Gaia data avalanche to throughly vet NYMG membership lists \cite{Mamajek2016}. Some groups might be merged, others split, and still others may be revealed to harbor distinct subgroups (e.g., \cite{Goldman2018}). Furthermore, the existence of a few NYMGs will likely be ``debunked,'' while a few new NYMGs will likely be discovered, through the application of cluster analysis to the combination of Gaia astrometry and sensitive, ground-based high-resolution optical/IR spectroscopy (\cite{LeeSong2018}; \S 3).\\
{\bf Dissipation into the field star population.} Gaia data, combined with ground-based high-resolution spectroscopic surveys that can establish RVs to $<$1 km s$^{-1}$ precision (e.g., \cite{Malo2014}), will provide the means to obtain  detailed kinematic evolutionary histories for NYMGs. These kinematic ``fingerprints'' will reveal the dynamical mechanisms through which, and timescales over which, members of young clusters and stellar associations eventually diffuse into the young field star population \cite{WrightMamajek2018,Binks2019,Roser2019,Tang2019}.

\vspace{-.08in}
\subsection{Astrophysics of young stars}    
\vspace{-.08in}

    
{\bf Age determination and refinement.} Completing the census of NYMGs and their members will allow stellar astronomers to obtain arguably the most elusive fundamental property of a star: its age. Precise age determinations for each NYMG will provide us with unique snapshots of different moments in stellar evolution, each for populations of several dozens to hundreds of low-mass stars. Ages of NYMGs as obtained from various methods --- e.g., color-magnitude diagrams (CMDs), Li depletion boundary, spectroscopic gravities, and traceback (convergent point) analysis --- can be assessed, compared, and calibrated, so as to evaluate the accuracy, applicability, and limitations of the age determination methods themselves \cite{MamajekBell2014,Goldman2018,Crundall2019}. Furthermore, thanks to their proximities and low line-of-sight extinction, NYMG CMDs yield perhaps the best empirical isochrones for ages $\sim$10--150 Myr \cite{Bell2015}, a range particularly poorly sampled by clusters in the solar neighborhood. In the Gaia era, nearby young associations thereby provide essential benchmarks for pre-main sequence evolutionary models \cite{Gagne2018VCA}. \\
{\bf Angular momentum evolution.} Over the next decade, rotational periods and velocities ($v \sin{i}$) can be obtained for most NYMG members from various photometric surveys (e.g., ASAS, TESS, SuperWASP, and soon, LSST) and high-resolution spectroscopy (\S 3), respectively. Hence, the early temporal evolution of stellar angular momentum as a function of mass can be ascertained, from several $M_\odot$ down to the H-burning limit. Among other things, these results will provide crucial insight into the  ($T_{eff}$) boundary between partially and fully convective young, low-mass stars. In turn, this will improve our understanding of turbulent dynamo theory (see next).\\
{\bf  Late-type star magnetic fields and high-energy radiation.} The combination of the rapid rotation of young, late-type stars and their large convection zones results in strong internal magnetic dynamos and, hence, large surface magnetic fields, which in turn manifest themselves externally in the form of intense high-energy (UV and X-ray) emission. 
The late-type members of NYMGs can provide especially keen insight into a crucial epoch in the early ($\sim$10--100 Myr) evolution of planetary systems during which remnant planet-forming disks and newborn planets are exposed to particularly hostile circumstellar radiation environments. 
The proximity of NYMGs enables both statistical studies of stellar high-energy radiation for coeval ensembles of stars (e.g., \cite{Kastner2016a}) as well as high-resolution UV and X-ray spectroscopy of individual stars (e.g., \cite{Kastner2004}) aimed at understanding such magnetically-driven chromospheric and coronal emission. 

\vspace{-.15in}
\subsection{Detection and characterization of exoplanets}
\vspace{-.08in}

{\bf Direct imaging and spectroscopy of young exoplanets.} Because young gas giant exoplanets are self-luminous in the infrared, young stars near Earth --- i.e., NYMG members --- offer by far the best targets for direct imaging and followup spectroscopy of giant exoplanets. Prime examples are the discoveries of young, ``warm,'' gas giant exoplanets orbiting HR 8799 \cite{Marois2008}, $\beta$ Pic \cite{LaGrange2010}, and 51 Eri \cite{Macintosh2015} (all of these stars are also orbited by debris disks; see next). Indeed, even with the advent of long temporal baseline precision astrometry and extremely large telescopes (ELTs) over the next decade, direct thermal IR imaging of nearby, young stars will continue to represent the {\it only} practical means available to detect exoplanets in wide ($\stackrel{>}{\sim}$20 AU, i.e., $\stackrel{>}{\sim}$100-year) orbits around their host stars \cite{Chauvin2016}. Meanwhile, Gaia+WFIRST precision astrometry potentially will enable dynamical mass measurements of any directly imaged young giant planets at smaller ($\sim$2--10 au) orbital semimajor axes  \cite{Brandt2019}, providing unique constraints on models of early planetary evolution.\\
{\bf Circumstellar disks.} Where studies of exoplanet birth environments are concerned, target proximity optimizes the high angular resolution and exquisite sensitivity of, e.g., ALMA (sub)mm and extreme adaptive optics (EAO) infrared imaging systems operating on giant telescopes. This premium on target proximity explains why, for example, the iconic, nearby classical T Tauri star TW Hya ($D = 60$ pc, age $\sim$8 Myr) has already been the subject of dozens of ALMA science programs (e.g., \cite{Bergin2016,HilyBlant2017,Huang2018}). Collectively, these radio interferometry and EAO programs targeting the many dozens of protoplanetary and debris disks orbiting NYMG member stars are deepening our understanding of all aspects of disk structure, composition, and evolution  (e.g., \cite{Rapson2015a, vanBoekel2017,Guzman2017,Bergner2018, Hendler2018,Kastner2018disk,Kennedy2018,Ribas2018}). Development of next-generation instrumentation for ALMA as well as an ALMA-like facility in the cm-wave regime \cite{VLAscibook} will further tap this enormous potential.  \\
{\bf Low-mass stars as exoplanet hosts.} The exoplanet community is now intensely focused on the characterization of low-mass (late K and M) stars, because of their large number and their favorable properties for detection of Earth-like, habitable exoplanets --- a prime science goal for future high contrast imaging and spectroscopy instruments on ELTs. 
The comprehensive characterizations of mass, radius, age, rotation, and magnetic activity for the statistically significant, coeval populations of late K and M stars that dominate NYMGs will be particularly significant in the context of improving our understanding of low-mass stars as habitable exoplanet hosts (see, e.g., \cite{Ramirez2014,Drake2019}).

\section{Required Key Advances within the Next Decade}
\vspace{-.08in}

Table 1 lists the key advances required to make progress in the foregoing ensemble of science themes in the 2020's. Here, we describe a few of these requirements 
in more detail. 

{\bf High-contrast thermal IR imaging:} 
At typical NYMG member distances ($\sim$50 pc), a coronagraph inner working angle of order 30 mas --- obtainable on 30-meter-class telescopes in the thermal IR --- corresponds to an orbital semimajor axis of 1.5 au. 
When used in (angular and/or reference star) differential imaging modes on an ELT, even present-day (e.g., vortex)  coronagraphic thermal IR imaging systems would easily deliver star-planet contrast ratios sufficient to detect a newly formed Jupiter-mass planet located at 5 au from a $\sim$10 Myr-old $\sim$1 $M_\odot$ star, or to detect Neptune-mass protoplanetary cores accreting disk gas at rates as small as $10^{-8}$ $M_J$ yr$^{-1}$, in integration times of an hour or less (scaling from results presented in \cite{Ruane2017}). By efficiently targeting all NYMG members within $\sim$120 pc with such high contrast imaging at sufficient sensitivity to detect substellar companions (e.g., \cite{Salama2018}), we will meet the parallel goal to obtain complete mass functions and multiplicity statistics for large coeval ensembles of stars. 

{\bf High-resolution optical/IR spectroscopy:} The high-resolution spectroscopic surveys that will be necessary to establish the ages and RVs (hence Galactic $UVW$ velocities) of the thousands of nearby young star candidates that have been and will be gleaned from Gaia data analysis must be sufficiently sensitive down to the useful magnitude limit of Gaia astrometry and photometry, i.e., $G \sim 20$. Exploitation of key spectral diagnostics of stellar youth --- e.g., Li absorption lines, features sensitive to surface gravities, emission lines indicative of stellar activity and accretion rate, $v \sin{i}$  --- requires optical/IR spectra with resolution $R>10^4$. Progress over the next decade is hence reliant on continued support for high-resolution spectrographic facilities on the largest (10- to 30-meter class) telescopes, so as to enable $R\sim10^4$ to $10^5$ near-IR spectroscopy at S/N $>$ 10 for stars with $G \sim 20$ in reasonable ($\sim$1 hr) exposure times. Such facilities will also enable precision RV searches for giant exoplanets within a few au of brighter NYMG members, thereby shedding light on the migration history of giant planets. 

{\bf High-resolution, high-throughput UV and X-ray spectroscopy:} Future UV and X-ray spectroscopic surveys of large samples of NYMG members can elucidate the waning stages of stellar magnetospheric accretion and the early evolution of stellar magnetic activity over the crucial epoch when planets and planetary atmospheres are just emerging from circumstellar, protoplanetary disks. Such surveys will be aimed at measurements of spectral line ratios diagnostic of plasma temperatures, densities, and abundances  (e.g., \cite{Brickhouse2010}). We are hence reliant on support for the development of (a) future X-ray satellite observatories, featuring micro-calorimeters or dispersive optics, that can achieve collecting area on the order of $10^4$ cm$^2$ \cite{Drake2019}, and (b) future space-based UV facilities that can replace (and far outperform) HST.

{\bf AI-assisted classification and cluster analysis algorithms:}  For all science themes listed in Table 1, it is imperative to establish a set of unified NYMG membership lists over the coming decade.  To avoid biases, this NYMG member list vetting process will be dependent on further development of unbiased pattern recognition algorithms, such as cluster analysis aided by supervised machine learning. Most cluster analyses of stellar groups have thus far focused on exploiting Galactic kinematics \cite{CastroGinard2018,CantatGaudin2018,Tang2019,Roser2019}, but we must eventually fold more complex and nuanced data (e.g., spectroscopic age diagnostics) into classification and clustering algorithms.

%% file: Table.tex
\vspace{-.08in}
\begin{table}[ht]
    \centering
    \caption{\sc Studies of Nearby, Young Stars: Drivers of Key Advances  in the 2020's}
    \begin{tabular}{c|c}
\hline
\hline
{\bf Science Goal} & {\bf Resources \& Key Advances Required} \\
\hline
\underline{\sc Young Star Populations }\\
mass functions, multiplicity &   30 mas imaging spectroscopy at $\ge$$10^8$ contrast \\
NYMGs: merge/split/discover/discard    & cluster analysis; neural networks \\
NYMG dissipation       &       high-res optical/IR spectroscopy to $G \sim 20$ \\
\hline
\underline{\sc Astrophysics of Young Stars}\\
Age determination &     high-res optical/IR spectroscopy to $G \sim 20$\\
Rotational evolution &  hourly-cadence photometry to $G \sim 20$ \\
High-energy radiation & high-$A_{eff}$, high-res UV \& X-ray spectroscopy \\
\hline
\underline{\sc Exoplanet Detection/Environments}\\
Direct imaging \& spectroscopy  &  30 mas imaging spectroscopy at $\ge$$10^8$ contrast\\
Protoplanetary \& debris disk studies &  5 mas resolution radio interferometry \\
 Exoplanet host characterization  &    high-res optical/IR, UV, X-ray spectroscopy \\
\hline
    \end{tabular}
    \label{tab:Requirements}
\end{table}

%% file: RefList2.tex



%% file: NYMGastro2020.bbl
\begin{thebibliography}{}

\bibitem{Bell2015}
{Bell}, C.~P.~M., {Mamajek}, E.~E., \& {Naylor}, T. 2015, \mnras, 454, 593.

\bibitem[{{Bergin} {et~al.}(2016){Bergin}, {Du}, {Cleeves}, {Blake}, {Schwarz},
  {Visser}, \& {Zhang}}]{Bergin2016}
{Bergin}, E.~A., {Du}, F., {Cleeves}, L.~I., {Blake}, G.~A., {Schwarz}, K.,
  {Visser}, R., \& {Zhang}, K. 2016, \apj, 831, 101

\bibitem[{{Bergner} {et~al.}(2018){Bergner}, {Guzm{\'a}n}, {{\"O}berg},
  {Loomis}, \& {Pegues}}]{Bergner2018}
{Bergner}, J.~B., {Guzm{\'a}n}, V.~G., {{\"O}berg}, K.~I., {Loomis}, R.~A., \&
  {Pegues}, J. 2018, \apj, 857, 69

\bibitem{Binks2019} Binks, A., Chalifour, M., Kastner, J., et al.\
  2019, \mnras, submitted 

\bibitem{Brandt2019} Brandt, T., et
  al. 2019, ``Realizing the Promise of High-Contrast Imaging:
Hundreds of Planets with Masses, Orbits, and Spectra
Enabled by Gaia+WFIRST-WFI Astrometry,'' Astro2020 Science White Paper

\bibitem{Brickhouse2010} Brickhouse, N. et al. 2010, ApJ, 710, 1835

\bibitem{CantatGaudin2018} Cantat-Gaudin, T., Jordi, C., Vallenari, A., et al.\ 2018, \aap, 618, A93 

\bibitem{CastroGinard2018} Castro-Ginard, A., Jordi, C., Luri, X., et al.\ 2018, \aap, 618, A59 

\bibitem{Chauvin2016}
{Chauvin}, G. 2016, in \emph{Young Stars \& Planets Near the Sun}, 
  J.~H. {Kastner}, B.~{Stelzer}, \& S.~A. {Metchev}, eds., Proc.\ of IAU Symposium
  314, pp. 213-219.

\bibitem{Crundall2019} Crundall, T., et al. 2019, arxiv:1902.07732 

\bibitem{Drake2019} Drake, J., et
  al. 2019, ``High-Energy Photon and Particle Effects on Exoplanet
  Atmospheres and Habitability,'' Astro2020 Science White Paper

\bibitem{ElBadry2018} El-Badry, K., \& Rix, H.-W.\ 2018, \mnras, 480, 4884 

\bibitem[Faherty et al.(2018)]{Faherty2018} Faherty, J.~K., Bochanski, J.~J., Gagn{\'e}, J., et al.\ 2018, \apj, 863, 91 

\bibitem{Faherty2019} Faherty, J.~K., et
  al. 2019, ``Brown Dwarfs and Directly Imaged Exoplanets in Young
  Associations,'' Astro2020 Science White Paper

\bibitem{Gagne2018BAN} Gagn{\'e}, J., \& Faherty, J.~K.\ 2018, \apj, 862, 138 

\bibitem{Gagne2018VCA} Gagn{\'e}, J., Faherty, J.~K., \& Mamajek, E.~E.\ 2018, \apj, 865, 136 

\bibitem[Gagn{\'e} et al.(2018)]{Gagne2018RNAAS} Gagn{\'e}, J., Gonzales, E.~C., \& Faherty, J.~K.\ 2018, Research Notes of the American Astronomical Society, 2, 17 

\bibitem[Gagn{\'e} et al.(2018)]{Gagne2018DR1} Gagn{\'e}, J., Roy-Loubier, O., Faherty, J.~K., Doyon, R., \& Malo, L.\ 2018, \apj, 860, 43 

\bibitem{Gagne2018a}
{Gagn{\'e}}, J., {Mamajek}, E.~E., {Malo}, L., {Riedel}, A., {Rodriguez}, D.,
  \emph{et~al.} 2018, \apj, 856, 23.


\bibitem{Gaia2018}
{Gaia Collaboration}, {Brown}, A.~G.~A., {Vallenari}, A., {Prusti}, T., {de
  Bruijne}, J.~H.~J., \emph{et~al.} 2018, ArXiv e-prints.

\bibitem{Goldman2018} Goldman, B., R{\"o}ser, S., Schilbach, E., Mo{\'o}r, A.~C., \& Henning, T.\ 2018, \apj, 868, 32 

\bibitem[{{Guzm{\'a}n} {et~al.}(2017){Guzm{\'a}n}, {{\"O}berg}, {Huang},
  {Loomis}, \& {Qi}}]{Guzman2017}
{Guzm{\'a}n}, V.~V., {{\"O}berg}, K.~I., {Huang}, J., {Loomis}, R., \& {Qi}, C.
  2017, \apj, 836, 30

\bibitem[Hendler et al.(2018)]{Hendler2018} Hendler, N.~P., Pinilla, P., Pascucci, I., et al.\ 2018, \mnras, 475, L62.


\bibitem[{{Hily-Blant} {et~al.}(2017){Hily-Blant}, {Magalhaes}, {Kastner},
  {Faure}, {Forveille}, \& {Qi}}]{HilyBlant2017}
{Hily-Blant}, P., {Magalhaes}, V., {Kastner}, J., {Faure}, A., {Forveille}, T.,
  \& {Qi}, C. 2017, \aap, 603, L6

\bibitem[{{Huang} {et~al.}(2018){Huang}, {Andrews}, {Cleeves}, {{\"O}berg},
  {Wilner}, {Bai}, {Birnstiel}, {Carpenter}, {Hughes}, {Isella}, {P{\'e}rez},
  {Ricci}, \& {Zhu}}]{Huang2018}
Huang, J., et~al. 2018, \apj, 852, 122

\bibitem{Kastner2004} Kastner, J.~H., et al. 2004, ApJ, 605, L49

\bibitem{Kastner2016a}
{Kastner}, J.~H., {Principe}, D.~A., {Punzi}, K., {Stelzer}, B., {Gorti}, U.,
  \emph{et~al.} 2016, \aj, 152, 3.

\bibitem{Kastner2016IAUSed} {Kastner}, J.~H., {Stelzer}, B., \& {Metchev}, S.~A. (eds.) 2016, \emph{{Young Stars \& Planets Near the Sun}}, Proceedings of IAU Symposium 314 (London: Cambridge U. Press).

\bibitem[Kastner et al.(2018)]{Kastner2018disk} Kastner, J.~H., Qi, C., Dickson-Vandervelde, D.~A., et al.\ 2018, \apj, 863, 106 

\bibitem[Kastner(2018)]{Kastner2018RNAAS} Kastner, J.~H.\ 2018, Research Notes of the American Astronomical Society, 2, 137 

\bibitem[Kennedy et al.(2018)]{Kennedy2018} Kennedy, G.~M., Marino, S., Matr{\`a}, L., et al.\ 2018, \mnras, 475, 4924.

\bibitem{LaGrange2010} Lagrange, A.-M., et al. 2010, Science,
329, 57

\bibitem{LeeSong2018} Lee, J., \& Song, I.\ 2018, \mnras, 475, 2955 

\bibitem{LeeSong2019} Lee, J., \& Song, I.\ 2019, \mnras, submitted

\bibitem[Macintosh et al.(2015)]{Macintosh2015} Macintosh, B., Graham, J.~R., Barman, T., et al.\ 2015, Science, 350, 64 

\bibitem{Malkov2018} Malkov, O., Karchevsky, A., Kaygorodov, P., Kovaleva, D., \& Skvortsov, N.\ 2018, arXiv:1811.04100 

\bibitem[Malo et al.(2014)]{Malo2014} Malo, L., Artigau, {\'E}., Doyon, R., et al.\ 2014, \apj, 788, 81 

\bibitem{MamajekBell2014} Mamajek, E.~E., \& Bell, C.~P.~M.\ 2014, \mnras, 445, 2169 

\bibitem{Mamajek2016}
{Mamajek}, E.~E. 2016, in \emph{Young Stars \& Planets Near the Sun}, 
  J.~H. {Kastner}, B.~{Stelzer}, \& S.~A. {Metchev}, eds., Proc.\ of IAU Symposium
  314, pp. 21--26.

\bibitem[Marois et al.(2008)]{Marois2008} Marois, C., Macintosh, B., Barman, T., et al.\ 2008, Science, 322, 1348 

\bibitem{VLAscibook} Murphy, E., et al., eds. 2018, {\it Science with a Next
    Generation Very Large Array}, ASP Conf.\ Ser., Vol.\ 517

\bibitem[Oh et al.(2017)]{Oh2017} Oh, S., Price-Whelan, A.~M., Hogg, D.~W., Morton, T.~D., \& Spergel, D.~N.\ 2017, \aj, 153, 257 

\bibitem[Ramirez \& Kaltenegger(2014)]{Ramirez2014} Ramirez, R.~M., \& Kaltenegger, L.\ 2014, \apjl, 797, L25 

\bibitem[{{Rapson} {et~al.}(2015{\natexlab{a}}){Rapson}, {Kastner}, {Andrews},
  {Hines}, {Macintosh}, {Millar-Blanchaer}, \& {Tamura}}]{Rapson2015a}
{Rapson}, V.~A., {Kastner}, J.~H., {Andrews}, S.~M., {Hines}, D.~C.,
  {Macintosh}, B., {Millar-Blanchaer}, M., \& {Tamura}, M. 2015{\natexlab{a}},
  \apjl, 803, L10


\bibitem[Ribas et al.(2018)]{Ribas2018} Ribas, {\'A}., Mac{\'\i}as, E., Espaillat, C.~C., et al.\ 2018, \apj, 865, 77.

\bibitem[R{\"o}ser et al.(2019)]{Roser2019} R{\"o}ser, S., Schilbach, E., \& Goldman, B.\ 2019, \aap, 621, L2 

\bibitem[Ruane et al.(2017)]{Ruane2017} Ruane, G., Mawet, D., Kastner, J., et al.\ 2017, \aj, 154, 73.

\bibitem{Salama2018} Salama, M., Ou, J., Baranec,
  C., et al.\ 2018, Adaptive Optics Systems VI, Proc.\ of the
  SPIE, 10703, 1070307 

\bibitem[Tang et al.(2019)]{Tang2019} Tang, S.-Y., Pang, X., Yuan, Z., et al.\ 2019, arXiv:1902.01404 

\bibitem{Torres2008}
{Torres}, C.~A.~O., {Quast}, G.~R., {Melo}, C.~H.~F., \& {Sterzik}, M.~F. 2008, in
  \emph{{Handbook of Star Forming Regions}}, ed. B. Reipurth, p. 757.

\bibitem[van Boekel et al.(2017)]{vanBoekel2017} van Boekel, R., Henning, T., Menu, J., et al.\ 2017, \apj, 837, 132.

\bibitem{WrightMamajek2018}
{Wright}, N.~J. \& {Mamajek}, E.~E. 2018, \mnras, 476, 381.

\bibitem[Zuckerman(2019)]{Zuckerman2019} Zuckerman, B.\ 2019, \apj, 870, 27.

\bibitem{ZuckermanSong2004}
{Zuckerman}, B. \& {Song}, I. 2004, \araa, 42, 685.

\end{thebibliography}
